\documentclass[12pt]{elsart}
\usepackage{graphicx,cite}
\usepackage{amsmath,amssymb}
\usepackage{upgreek}
\usepackage{color}

\begin{document}

\begin{frontmatter}
\title{\mathversion{bold} A Search for Hidden Photon CDM in a Multi-Cathode Counter (MCC) data}

\author{A.~Kopylov},
\author{I.~Orekhov},
\author{V.~Petukhov}

\address{Institute for Nuclear Research of RAS, Prospect of 60th Anniversary of October Revolution
7A, Moscow, Russia}

\begin{abstract}
%% Text of abstract
%%
A search for hidden photon cold dark matter in a mass range from 5
to 500 eV in data collected during 60 days in November and December,
2015 by a Multi-Cathode Counter (MCC) is reported. From the analyses
of this data we found no evidence for the existence of HP CDM and
set an upper limit on the photon-HP mixing parameter $\chi $. This
is the first result obtained by direct measurements in this mass
range for hidden photon CDM using a single electron event in MCC as
a signature.
\end{abstract}


\end{frontmatter}
\clearpage
%\begin{linenumbers}
\def\spreadlines#1{\par\renewcommand\baselinestretch{#1}\normalsize}
\def\red{\color{red}}  \def\blue{\color{blue}}

\section{Introduction}\label{Introduction}

Hidden-sector photons as massive particles having only a tiny
interaction (kinetic mixing) with known particles was suggested in
\cite{Okun} where it has been also emphasized that stellar evolution
can place a strong limit for these species. However recently the
interest to HPs as the candidate for CDM has been revived probably
because the experiments on WIMPS as a paradigm for dark matter
produce negative results what stimulates researchers to look for new
horizons.

The low energy Lagrangian of this model

\begin{equation}
{\mathcal{L}=-\frac{1}{4}F_{\mu\nu}F^{\mu\nu}-\frac{1}{4}{X}_{\mu\nu}{X}^{\mu\nu}-
\frac{\chi}{2}F_{\mu\nu}{X}^{\mu\nu}+\frac{m_{\gamma'}^{2}}{2}{X}_{\mu}{X}^{\mu}+J^{\mu}A_{\mu}
\label{eq.1}}
\end{equation}

where $F_{\mu\nu}$ is the field strength of the ordinary
electromagnetic field $A^\mu$, and $X_{\mu\nu}$ is the field
strength of the HP field $X^\mu$. Lagrangian contains a mixing term
with $\chi$ - the mixing parameter for a given mass $m_{\gamma '}$
of a HP \cite{Holdom} which determines the probability of HP-photon
conversion by which HP can be ``seen'' in experiment. Recently the
eV mass range of HP CDM was investigated with a dish antenna
\cite{Suzuki} a novel method proposed in \cite{Horns}. The method
suggests detection of a reflected from metallic mirror (antenna)
electromagnetic wave which is emitted by the oscillation of
electrons of the antenna's surface induced by the tiny electric
field of HP. Noteworthy, the target in this novel method is not a
mass but a surface of the detector because the conversion of HP into
photon occurs at the interface of metal-dielectric. The method
should work only if the reflectance of mirror is high. But what
would be if the reflectance of mirror is low? Here in our work we
made a focus exactly on this case, assuming that at higher masses of
HPs the oscillation of electrons of antennas surface induced by HPs
will produce with certain quantum efficiency the emission of single
electrons from metal. We take this value equal to the quantum
efficiency for real photon with energy $\omega = m_{\gamma'}$ to
emit electron from the surface of a metal. Thus we proposed as a
method for detection of hidden photon CDM to search for the events
with single-electron emission from metallic surface as a signature
of the conversion HP-photon. To record these events a detector
should be highly sensitive to single electrons emitted from the
surface and also should be devised a way for subtraction of the
background from other sources what is very essential part of the
work. As a special technique to solve this task we devised a
Multi-Cathode Counter (MCC) which has been designed, assembled and
tested during last year. This counter has been described in detail
in \cite{Kopylov1}, \cite{Kopylov2}. We report here the result
obtained from data collected during 60 days in November and
December, 2015 on this multi-cathode counter.

\section{Experimental part}\label{Experimental part}

\begin{figure}[htb]
\begin{center}\vspace{-0.2cm}
\includegraphics[width=8cm]{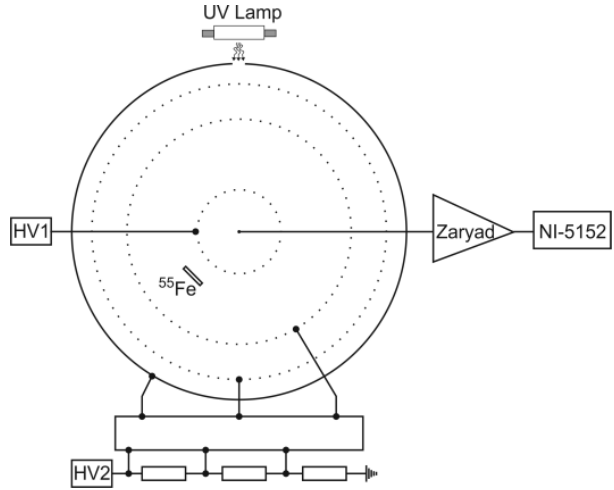}
\caption{Simplified electronic scheme.} \label{Figure 1}
\end{center}
\end{figure}

The general view of the counter can be found in \cite{Kopylov1}, the
simplified electronic scheme is presented on Fig.~\ref{Figure 1}. It
is a gaseous proportional counter filled with argon-methane (10\%)
mixture at 0.2 MPa with four cathodes. The counter has a central
anode wire of 20 mm, three cathodes made of 100 mm nichrome wires
tensed with a pitch of a few mm and a fourth metal cathode which
acts as ``antenna'' for HPs. The external ($4^{th}$) cathode of the
counter made of copper cylinder has 194 mm in diameter and 400 mm in
length. The cathode has been cleaned by a routine technique.
Commercially available copper was chemically etched, rinsed with
distilled water and ethanol. It has relatively large ($\approx 20.2$
$m^2$) surface which acts as a ``mirror''. The detector is counting
electrons emitted from a copper cathode under the impact of hidden
photons with $\omega = m_{\gamma'}\approx 5\div500$ eV. Because the
reflectance of the ``mirror'' is low at these energies it is assumed
that one should observe not photons reflected from the ``mirror''
but electrons emitted from a copper cathode.

The counter is used in three different configurations. In first
configuration used to measure the count rate $R_1$ of single
electrons emitted from a copper cathode electrons emitted from
copper drift freely into the central section with high gas
amplification. In $2^{nd}$ configuration electrons emitted from
copper are scattered back in argon at 0.2 MPa by higher negative
potential of $3^{rd}$ cathode which acts as a barrier. In $3^{rd}$
configuration the highest negative potential is applied to $2^{nd}$
cathode which acts as a barrier. As a measure of the effect from
hidden photons we use the expression

\begin{equation}
R_{MCC}=R_1-(R_2-D_3/D_2 \cdot R_3) \label{eq.2}
\end{equation}

here $R_2$ ($R_3$ ) is the count rate in $2^{nd}$ ($3^{rd}$)
configurations, $D_3$ ($D_2$) is a diameter of the $3^{rd}$
($2^{nd}$) cathode. This expression enables to subtract the
background from the ends of the counter with distorted electric
fields and also the one from single electrons drifted from a gas
phase outside of the cathode in $2^{nd}$ configuration as it was
explained with all details of counting procedure in \cite{Kopylov2}.
The counter has been placed in a steel cabinet with 30 cm iron
shield. The count rate of single-electron events decreased by a
factor of 2 in comparison with the one when detector was outside of
the shield while the flux of gammas in the region between peaks 511
and 661.6 keV was attenuated by a factor of 30. For further
reduction of background count rates it would be highly desirable to
collect data in an underground chamber where the flux of muons is
decreased by many orders of magnitude. This part of the work is in
our plans also.

\begin{figure}[htb]
\begin{center}\vspace{-0.2cm}
\includegraphics[width=8cm]{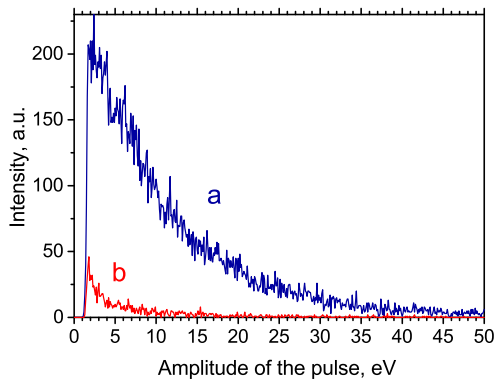}
\caption{The single electron spectra obtained in measurements in
$1^{st}$ (a) and $2^{nd}$ (b) configurations at the same flux of UV
photons. The conversion factor is $\approx$ 2.27 eV/mV.}
\label{Figure 2}
\end{center}
\end{figure}

For calibration of the counter we used $^{55}Fe$ source and UV light
of the mercury lamp.  High voltages in all three configurations were
picked up to get a gas amplification of about $10^5$ so that
amplitude of 5.9 keV peak on the output of charge sensitive
preamplifier reached a level 1400 mV. The counter was working in a
mode of limited proportionality which could be observed by two peaks
of $^{55}Fe$ source as it was explained in \cite{Kopylov2}. Then the
internal walls of the counter were irradiated by UV light of a
mercury lamp through a window made of melted silica.
Fig.~\ref{Figure 2} shows the single electron spectra obtained in
measurements in $1^{st}$ and $2^{nd}$ configurations by the same
intensity of UV light.

One can see that in the $2^{nd}$ configuration the count rate was
about 10 times lower than in the first one. It means that electrons
emitted from the walls of the counter in $2^{nd}$ configuration were
scattered back and did not reach a central section. Thus a count
rate in $2^{nd}$ configuration is a background and should be
subtracted. Single electron spectra obtained in all 3 configurations
had an exponential shape with the same inverse index of exponent in
the range $8.4\pm0.4$ mV. By the conversion factor 2.27 eV/mV it is
near to the average energy 27 eV to produce electron-ion pare in
argon. Data were collected by frames. Each frame contained 2M
points, each point 100 ns. The collected data were stored on a disk
then the collection resumed. The analysis of the collected data was
done off-line. The noisy periods were removed from analysis.

\section{Sensitivity of the method}\label{Sensitivity of the method}

By conducting this experiment it was assumed that if hidden photons
have a mass greater than a work function of a metal of the cathode
then passing through a cathode they will induce the oscillation of
the current on the surface of a cathode which will stimulate
emission of electrons from metal. The very substantial point is that
the conversion of HPs into electrons emitted from the metal occurs
at the interface metal -- dielectric. The counter detects fields,
not particles, so the effect from HP is proportional to the surface
of the cathode. As it was suggested in \cite{Horns} for antenna,
here it was assumed that if DM is totally made up of hidden photons,
the power concentrated on the cathode of the counter is

\begin{equation}
{P=2\alpha^{2}\chi^{2}\rho_{CDM}A_{MCC} \label{eq.3}}
\end{equation}

where $A_{\rm MCC} = 0.2$ $m^2$ is the surface of the cathode of the
counter and $\alpha=|cos(\theta)|$  if HPs are oriented in the same
direction but $\alpha=\sqrt{2/3}$ if HPs have random orientation.
For our detector we should take for P the value
$R_{MCC}m_{\gamma'}/\eta$. Then from the expression Eq.(\ref{eq.3})
taking $\alpha=\sqrt{2/3}$ one can easily obtain sensitivity:

\begin{equation}
\begin{split}
\chi_{{\rm
sens}}=\end{split}2.9\cdot10^{-12}\left(\frac{R_{MCC}}{\eta\,
1\,{\rm Hz}}\right)^{\frac{1}{2}}\left(\frac{m_{\gamma'}}{{\rm
eV}}\right)^{\frac{1}{2}}\left(\frac{0.3\,{\rm GeV/cm^3}}{\rho_{{\rm
CDM,\, halo}}}\right)^{\frac{1}{2}}\left(\frac{1\,{\rm
m^{2}}}{A_{{\rm
MCC}}}\right)^{\frac{1}{2}}\left(\frac{\sqrt{2/3}}{\alpha}\right)
\label{eq.4}
\end{equation}

\section{The results obtained from data collected on MCC}\label{The results obtained from data collected on MCC}

\begin{figure}[htb]
\begin{center}\vspace{-0.2cm}
\includegraphics[width=8cm]{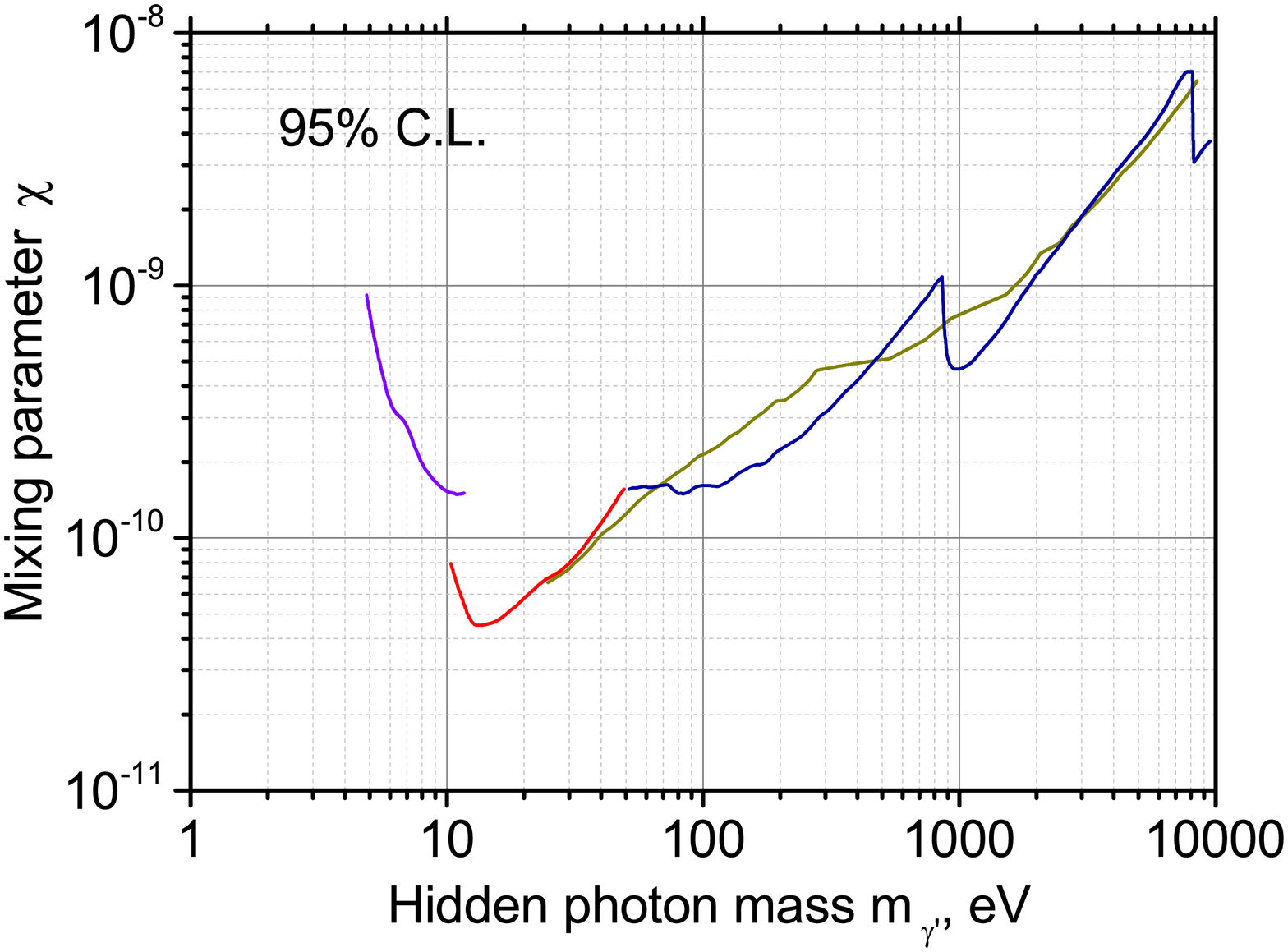}
\caption{The upper limit for a mixing constant $\chi $.}
\label{Figure 3}
\end{center}
\end{figure}

We report results from data collected during 60 days of the search
for HPs in November and December, 2015. The scattering of the
experimental points $R_1$, $R_2$, $R_3$ has been used for the
evaluation of the total uncertainty of measurements. The average
value of $R_{MCC}$ calculated for this period was found to be
$\bar{R}_{MCC}=0.06\pm0.18$ Hz. If to take the normal distribution
for uncertainties then we obtain that at 95\% confidence level:
$R_{MCC}<0.42$ Hz. We found no evidence of the existence of HP CDM.
Fig.~\ref{Figure 3} shows the upper limit obtained from the
expression Eq.(\ref{eq.4}) as a function of the mass of the hidden
photon. Here for masses of HPs $m_{\gamma'}<$ 11.6 eV (magenta) the
quantum efficiency $\eta$ was taken from \cite{Krolikowski}, for 10
eV $< m_{\gamma'} <$ 60 eV (red) -- from \cite{Cairns}, for
$m_{\gamma'}$ from 20 eV till 10 keV (green) -- from \cite{Day} and
from 50 eV till 10 keV (blue)-- from \cite{Henneken}.

One can see that the highest sensitivity has been reached in the
interval from approximately 10 eV till about 20 eV (red line). The
major source of systematic error is the uncertainty in quantum
efficiency. The estimated uncertainty in determination of $\eta $
for copper in this range of masses is about $\pm 30\%$
\cite{Cairns}. The difference between green and blue line is
explained by difference in cleaning procedure of copper. Green line
- when quantum efficiencies are taken for routinely cleaned copper
(by solvents) while blue line - for meticulously done preparation of
copper specimen by evaporation in high vacuum without any contact
with atmosphere. For meticulously cleaned copper the impact of
electron shells is clearly seen in the data while for routinely
cleaned copper the effect of electron shells is rather smoothed.
This can explain also a divergence at 10 eV (magenta -- carefully
prepared clean Cu films while red -- routinely cleaned copper).

\section{Conclusion}\label{Conclusion}

A new technique of Multi Cathode Counter (MCC) has been developed to
search for hidden photon CDM by single electrons emitted from the
surface of metal.  The sensitivity was estimated in the assumption
that all dark matter is composed of hidden photons (HP). In method
suggested it was assumed that HPs of the mass greater than a work
function of the metal, the cathode of the counter is fabricated
induce emission of single electrons from a cathode. The peculiarity
of this signature means that conversion of HP in photon occurs at
the interface metal-dielectric. The results have been obtained for
HPs with a mass from 5 to 500 eV from data collected during 60 days
in November and December, 2015. From the analysis of this data we
found no evidence for the existence of HP CDM and set an upper limit
on the photon-HP mixing parameter $\chi$ with a minimum on the level
of about $5\cdot10^{-11}$ at 95\% CL for the hidden photon mass
$12\div15$ eV. Stellar astrophysics provides stringent constraint
for this value. Our result is deep inside the regions excluded by
astrophysical models; see for example \cite{Haipeng} and references
therein. The new thing is that this result with HP CDM was obtained
in direct measurements with a very peculiar signature of conversion
HP-photon at the interface metal-dielectric. In our plans are to
continue the measurements to collect more data and to refine the
procedure of data treatment. At present time we are also
constructing a new detector with a more developed design and with
this new detector we are planning to collect data in an underground
laboratory with a very low flux of muons.

\section{Acknowledgments}

The authors appreciate very much the help provided by E.P.Petrov and
A.I.Egorov in fabrication of the counters and to Grant of Russian
Government "Leading Scientific Schools of Russia \#3110.2014.2" for
partial support of this work.

\end{document}